\documentstyle[fund]{article}

\renewcommand{\textwidth}{16.5cm}

\newcommand{\Ncollisions}{{\cal N}_{\small coll}}

\newcommand{\Qstar}{Q^{*}}

\newcommand{\Rt}{{\cal R}_t}
\newcommand{\Rtdot}{\dot{\Rt}}

\newcommand{\Sdplusone}{S^{(d+1)}}
\newcommand{\Sone}{S^{(1)}}

\newcommand{\ta}{t_a}

\newcommand{\tstartwo}{t_{2}^{*}}
\newcommand{\tstarmany}{t_{m}^{*}}
\newcommand{\tl}{t_l}

\newcommand{\xt}{x_t}

\newcommand{\r}{{\bf r}}

\newcommand{\nainf}{n_{A}^{\infty}}
\newcommand{\nbinf}{n_{B}^{\infty}}

\newcommand{\na}{n_A}
\newcommand{\nb}{n_B}
\newcommand{\nas}{n_A^s}
\newcommand{\nbs}{n_B^s}

\newcommand{\rhoabs}{\rho _{AB}^{s}}
\newcommand{\rhoab}{\rho_{AB}}

\newcommand{\rhobab}{\rho_{BAB}}

\newcommand{\dc}{d_{\rm c}}

\newcounter{fignumber}

\begin{document}



\renewcommand{\thepage}{}

\titleben{\LARGE \bf Interfacial Reactions: Mixed Order Kinetics and
Segregation Effects}

\author{\Large 
BEN O'SHAUGHNESSY\ $^{1}$ \ and \ DIMITRIOS VAVYLONIS\ $^2$ \\ 
}

\maketitle

\ \\ \newline
{\large $^1$ Department of Chemical Engineering, Columbia University, New York, NY 10027} \\
\ \\
{\large $^2$ Department of Physics, Columbia University, New York, NY 10027} \\
\ \\ \ 



\large

\pagenumbering{arabic}

\section*{ABSTRACT}

We study A-B reaction kinetics at a fixed interface separating A and B
bulks.  Initially, the number of reactions ${\cal R}_t \sim t n_A^\infty
n_B^\infty$ is 2nd order in the far-field densities $n_A^\infty,n_B^\infty$.
First order kinetics, governed by diffusion from the dilute bulk,
onset at long times: ${\cal R}_t\approx x_t n_A^\infty$ where $x_t\sim t^{1/z}$
is the rms molecular displacement.  Below a critical dimension,
$d<d_c=z-1$, mean field theory is invalid: a new regime appears,
${\cal R}_t\sim x_t^{d+1} n_A^\infty n_B^\infty$, and long time A-B segregation
(similar to bulk $A+B\gt\emptyset$) leads to anomalous decay of
interfacial densities.  Numerical simulations for $z=2$ support the
theory.

\pagebreak


A considerable analytical and numerical research effort has addressed
the kinetics of bimolecular reactions in a bulk phase
\citeben{kotominkuzovkov:book_short,ovchinnikov:segregation_combo,%
bramson:aplusb_combo_fund,gennes:polreactionsiandii}.  These are
complex many body systems; correlation functions of different order
are coupled in an infinite hierarchy of dynamical equations
\citeben{doi:reaction_secondquant1and2}.  Analytical treatments have
employed decoupling approximations which allow truncation of the
hierarchy \citeben{kotominkuzovkov:book_short}, and more recently
renormalization group techniques
\citeben{lee:aplusb_rg_all_aip_prl}.  From these
studies it is known that the classical mean field (MF) theory is valid
only above a critical spatial dimension $d_c$.  According to MF
kinetics, the net reaction rate is simply proportional to a product of
spatially and thermally averaged densities.  For the single-species
case ($A+A \gt \emptyset$) $d_c=2$, while $d_c=4$ in the two-species
case ($A+B \gt \emptyset$).  In lower dimensions behavior is very
different.  For example in the two-species case, Ovchinnikov and
Zeldovich, and Toussaint and Wilczek
\citeben{ovchinnikov:segregation_combo} established a remarkable
segregation at long times into A-rich and B-rich domains; MF kinetics
break down and the asymptotic decay of density fields no longer
follows the $1/t$ MF prediction.  All of these findings concern
non-interacting small molecules, for which the rms diffusive
displacement after time $t$ follows Fick's law, $\xt\twid t^{1/2}$,
independently of spatial dimension $d$.  For systems with arbitrary
(dimension-independent) dynamical exponent $z$, $\xt\twid t^{1/z}$,
the generalizations are $d_c=z$ and $d_c=2z$ for $A+A\gt
\emptyset$ and $A+B\gt \emptyset$, respectively
\citeben{gennes:polreactionsiandii}.

In contrast to the bulk, little is understood theoretically about {\em
interfacial} reaction kinetics.  Unlike the bulk $A+B \gt \emptyset$
situation, the species A and B may now only react at a permanent
interface separating the bulk A and B phases (see fig.
\ref{iface_letter}).  Applications involving reactions of this type
include a large class where small molecules ($z=2$) react at
liquid-liquid, liquid-solid or solid-solid interfaces
\citeben{doraiswamysharma:book}.  In another important class,
functional groups attached to long polymer chains ($z=4,8$) react at
an interface separating immiscible polymer melts.  The A-B copolymers
formed by reactions stabilize and reinforce the interface
\citeben{scott:scottmacosko_gersappe:combo_aip}.  In these systems,
which are the subject of this letter, the two bulk phases are forever
separated by a permanent interface of fixed width.  A very different
but conceptually related class of systems, which has been addressed by
many works \citeben{galfi:aplusb_front_combo}, is that of {\em
non-stationary} reactive chemical fronts where the A and B bulk phases
mix and the interface broadens as reactions proceed.  Other more
distantly related models include catalytic
reactions on surfaces such as the ``monomer-monomer'' model
\citeben{meakinscalapino:catalysis_model}, $A + B \gt \emptyset$ with
spontaneous generation of particles \citeben{lindenberg:aplusb_input},
and reaction fronts near semipermeable walls
\citeben{chopard:aplusb_semipermeable}.

In this letter we present a theoretical study of interfacial reaction
kinetics
\citeben{ben:reactiface,benfred:reactiface_combo_aip}.  
Our principal findings are as follows.  (1) MF kinetics break down
below a critical dimension $d_c=z-1$.  (2) For spatial dimensions
$d<d_c$, a short time diffusion controlled (DC) regime occurs with the
number of reactions per unit area growing as $\Rt \approx \xt^{d+1}\,
\nainf \nbinf$, where $\nainf, \nbinf$ are the far field densities.
(3) For $d<\dc$, at long times reactants segregate into A-rich and
B-rich domains at the interface.  Correspondingly, interfacial
densities decay with non-mean-field power laws.  (4) Reaction kinetics
are of mixed order.  In all cases short time 2nd order kinetics cross
over at long times to kinetics which are 1st order in the density on
the {\em more dilute} A side: $\Rt\approx \xt\,\nainf$.

These results are derived without resorting to ad-hoc decoupling
approximations.  Instead, we postulate physically motivated bounds on
the correlation functions.  It is possible that these bounds might be
proved rigorously, but we do not attempt this here.  Having made these
assumptions, the subsequent analysis is exact.

Our principal aim is the reaction rate per unit area, $\Rtdot\equiv
d\Rt/dt$, proportional to the number of A-B pairs in contact at the
interface:
                                                \begin{eq}{rate}
\Rtdot = \lambda \rhoabs(t) \comma \gap \lambda \equiv Q h a^3 \period
                                                                \end{eq}
Here $Q$ is the local reactivity, $h$ is the interface width, and $a$
the reactive group size.  $\rhoabs$ is the 2-body correlation function
evaluated at the interface.  In addition, we seek the mean density
profiles on the A and B sides, $\na(\r), \nb(\r)$, whose
characteristic features are: the far field values, $\nainf, \nbinf$;
the values at the interface ($\r=0$), namely $\nas$ and $\nbs$; and
the size of the depletion region (if any) near the interface.  Using
Doi's \citeben{doi:reaction_secondquant1and2} second quantization
formalism for classical many-body reacting systems, we have derived
the following exact expression:
                                                \begin{eq}{fate}
\nas(t) = \nainf - \lambda \int_0^t dt' \, \Sone_{t-t'} \, \rhoabs(t')  \period
                                                                \end{eq}
Here $\Sone_t \approx 1/\xt$ is the one dimensional return
probability: the probability an A or B group, initially at the
interface, returns to it after time $t$ in the absence of reactions.
The integral term simply subtracts off A reactants which failed to
arrive at the interface at time $t$ due to earlier reactions.

The technical difficulty is already apparent.  The reaction rate and
interfacial densities involve the 2-body correlation function
$\rhoab$.  But one can show (see below) that the dynamics of $\rhoab$
involve 3-body correlation functions; these in turn are coupled to
4-body correlations, and so on.  This is the infinite hierarchy.  How
can one close eqs. \eqref{rate} and \eqref{fate}?  A simple way to
achieve this is to assume MF kinetics, \ie to neglect density
correlations at the interface:
                                                \begin{eq}{approximation}
\rhoabs(t) \approx \nas(t) \, \nbs(t)  \gap 
\mbox{(MF approximation)} \period
                                                                \end{eq}
The reaction rate is then simply proportional to the product of
interfacial densities.

Let us proceed by simply assuming MF kinetics are valid.  We return
later to the question of when this assumption breaks down.  Consider
first the symmetric case, $\nainf=\nbinf$.  Now since the integral
term in eq. \eqref{fate} is zero initially and grows continuously, at
short times it must be much less than $\nainf$, and hence $\nas = \nbs
\approx \nainf$.  Using the MF approximation, eq.
\eqref{approximation}, one sees that the integral term then increases
as $\lambda \nainf \nbinf \, t /\xt \twid t^{1-1/z}$ and thus becomes
of order $\nainf$ at a timescale
                                                \begin{eq}{tstarmany}
\tstarmany = \ta (\lambda\ta \nbinf/a)^{z/(1-z)} \comma
                                                                \end{eq}
where $\ta$ is the diffusion time corresponding to $a$.  Thus for
times greater than $\tstarmany$, $\nas$ tends to zero and the integral
term now balances with $\nainf$ in eq. \eqref{fate}.  Seeking a power
law solution for $\nas$, one immediately obtains its long time decay:
                                                \begin{eq}{snow}
\nas \approx \casesbracketsshortii{\nainf}                     {t \ll \tstarmany}
             {(\xt\, \nainf/ t \lambda)^{1/2} \twid\, t^{(1-z)/(2z)}}
                                                         {t \gg \tstarmany}
\gap (\mbox{MF}) \period
                                                                \end{eq}
The number of reactions, from eqs. \eqref{rate} and
\eqref{approximation}, is thus
                                                \begin{eq}{chair}
\Rt \approx \casesbracketsshortii{\lambda t \ \nainf \nbinf} 
                                        {t \ll \tstarmany}
                               {\xt \ \nainf \, \twid \, t^{1/z}}
                                        {t \gg \tstarmany}
\gap \mbox{(MF)} \period
                                                                \end{eq}
These reaction kinetics are rather novel: they are not of fixed order.
The short time 2nd order behavior crosses over to long-time 1st order
kinetics.

\ignore{
The asymmetric case is 
The difference in the asymmetric case, $\nbinf > \nainf$, is that now
$\nbs$ tends to a finite value.  Notice that an identical equation to
eq. \eqref{fate} holds for $\nbs$.  Subtracting these two equations
gives $\nbs - \nas = \nbinf -\nainf$.  Thus the long time solution to
eq. \eqref{fate} is now $\nbs \approx \nbinf -\nainf$ and $\nas
\approx (\nainf/\lambda)/(\nbinf- \nainf) \ d\xt/dt \twid t^{1/z-1}$.
} 

The above results have a very clear physical interpretation.  At short
times, interfacial densities are unchanged from their initial values.
But by time $t$ an A reactant initially within diffusive range of the
interface (\ie closer than $\xt$) will have collided with it of order
$(t/\ta) (h/\xt)$ times.  Each collision produces reaction with
probability $\approx \nbinf a^d Q\ta$.  By time $\tstarmany$,
therefore, the net reaction probability becomes of order unity.  Thus
for $t>\tstarmany$ a depletion hole of size $\xt$ grows at the
interface, the reaction rate is diffusion controlled and first order
kinetics onset.  The expression $\Rt\approx \xt \nainf$ is just the
total number of A molecules per unit area within $\xt$ of the
interface.  Equating its time derivative to the expression for
$\Rtdot$ implied by eqs. \eqref{rate} and \eqref{approximation}, one
immediately obtains the long-time decay of the interfacial density,
$\nas \twid t^{(1-z)/(2z)}$.

The analysis for the asymmetric case, $\nbinf > \nainf$, is similar
except that we find $\nbs$ asymptotes a finite value,
$\nbs(t\gt\infty) \approx \nbinf -\nainf$, while $\nas$ tends to zero.
The reaction kinetics of eq. \eqref{chair} are unchanged.  Physically,
this means a density hole of size $\xt$ grows on the more {\em dilute}
A side.  It is delivery of the A species to the interface which
determines $\Rt$.

When are these MF results valid?  To answer this question properly,
one must examine the dynamics of $\rhoabs$.  Using Doi's formalism
\citeben{doi:reaction_secondquant1and2} we have derived an exact
self-consistent relation for $\rhoabs$ which involves the 3-body
correlation $\rhobab(\r|0,0;t)$, namely the conditional density of B
groups at $\r$, given an A-B pair at the origin.  This relation reads
                                                \begin{eqarray}{tree}
\rhoabs(t') &=& \nainf \nbinf - 
                \lambda \int_0^t dt'\, \Sdplusone_{t-t'}\, \rhoabs(t') 
                - I_{BAB}(t) - I_{ABA}(t) ;       
                                        \drop
I_{BAB}(t) &\equiv& \lambda \int_0^t dt' \int d\r\ {\cal G}_{t-t'}(\r)\,
                              \rhobab(\r|0,0;t')\, \rhoabs(t')               
\comma
                                                     \end{eqarray}
where $I_{ABA}$ equals $I_{BAB}$ with A and B interchanged.  Here
$\Sdplusone_t \approx 1/\xt^{d+1}$ is the probability an A-B pair is
in contact at the interface at $t$, given its members were in contact
at the interface initially, in the absence of reactions.
${\cal G}_t(\r)$ is the probability a pair is in contact at the
interface at $t$, given initial pair separation $\r$ with one member
being at the interface.

Now eq. \eqref{tree} is not in a closed form for $\rhoabs$, since it
contains unknown 3-body terms.  We are able to close eq. \eqref{tree}
after postulating physically motivated bounds on the 3-body
correlation functions.  These are much weaker assumptions than those
involved in the typical procedure which entails approximating
$\rhobab$ as a product of lower order correlation functions.  We
postulate that there exist constants $U$ and $L$ of order unity such
that: (1) $\rhobab(\r|0,0;t) \le U\, \nbinf$ and (2)
$\rhobab(\r|0,0;t) \ge L\, \nbinf$ for $x>\xt$, where $x$ is the
distance from the interface.  Assumption (1) states our physical
expectation that conditional densities never become much greater than
the far-field densities.  Assumption (2) states that conditional
densities at points beyond diffusional range of the interface are
uncorrelated with it.

These assumptions immediately imply a maximum and a minimum value for
$\rhobab$ for each $\r$.  The maximum value is $U \nbinf$ for all
$\r$, while the minimum value is zero for $x<\xt$ and $L \nbinf$ for
$x>\xt$.  We can thus obtain bounds on $I_{BAB}$ by substituting these
two extreme cases into its definition, eq. \eqref{tree}.  The
important point is that these two bounds are of the same order; we
have thus specified $I_{BAB}$ to within a time-dependent prefactor of
order unity.  Doing the same for $I_{ABA}$, we find after substitution
in eq. \eqref{tree}
                                                \begin{eq}{door}
\rhoabs(t) = \nainf \nbinf - 
                \lambda \int_0^t dt'\, \Sdplusone_{t-t'}\, \rhoabs(t') 
              - \lambda n(t) \int_0^t dt'\, \Sone_{t-t'}\, \rhoabs(t') 
\comma
                                                                \end{eq}
where $n(t) \equiv A(t) \, [\nainf + \nbinf]$, and $A$ is a bounded
positive function of order unity.  The exact form of $n(t)$ is
unknown; however, we have found that the vanishing of the interfacial
density on the A side at long times implies $n(\infty)=\nbinf$
exactly.  Since the term involving $n(t)$ is relevant at long times
only, in effect $n$ may be replaced by $\nbinf$.

\ignore{
where $n(t) \equiv A(t) \, [\nainf + \nbinf]$, with $A$ being a
bounded positive function of order unity.  The exact form of $n(t)$ is
unknown; however, the term containing it is is relevant at long times
only and $n(t)$ can be replaced by $n(\infty)$ without big error.  We
were able to determine that $n(\infty)=\nbinf$ (B is the more dense
side) after making some further simple assumptions.  These amount to
assuming that at the interface, reactions may only induce positive
correlations between like particles and negative correlations between
A and B's.
} 

It is now straightforward to solve eq. \eqref{door} for $\rhoabs$, and
thus obtain the reaction rate via eq. \eqref{rate}.  One can show that
deletion of the term containing $\Sdplusone$ reproduces the MF
kinetics of eq. \eqref{chair}.  This term is indeed irrelevant above a
critical dimension, $d>\dc=z-1$.  It is also irrelevant for $d<\dc$ if
the reactivity $Q$ is smaller than a certain value, $Q<\Qstar$ (see
below).

For lower dimensions and high reactivities, however, we find that
during a certain interval $\tstartwo<t<\tl$ this same term, the term
containing $\Sdplusone$ in eq. \eqref{door}, is dominant.  Then the MF
approximation breaks down and reaction kinetics are of second order
and DC.  This is a new regime whose physical origin is as follows.
Consider an A and a B molecule which happen to be so close to each
other that their exploration volumes overlap by time $t$ (see fig.
\ref{iface_letter}).  
How many A-B collisions, $\Ncollisions$, have there been by time $t$?
The A molecule visited the interface of order $(t/\ta) (h/\xt)$ times,
and during each visit encountered the B molecule with probability
$(a/\xt)^d$.  Hence $\Ncollisions\approx (t/\ta) (ha^d/\xt^{d+1})$,
and the total reaction probability $Q\ta\Ncollisions\twid
t^{(\dc-d)/z}$ is thus an increasing function of time for $d<d_c$.  It
reaches unity at a time $\tstartwo$ where
                                                \begin{eq}{rupp}
\tstartwo = \ta (Q \ta h/a)^{z/(d-\dc)} \period
                                                                \end{eq}
Below the critical dimension, therefore, for $t>\tstartwo$ any A-B
pair with separation $\xt$ or less will definitely have reacted by
time $t$.  Thus a depletion hole develops in the 2-body correlation
function, invalidating the MF assumption.  Instead, $\Rt$ is
proportional to the number of such pairs per unit area, $\xt^{d+1}
\nainf \nbinf$.  The kinetic sequence is now
                                                \begin{eq}{paper}
\Rt \approx \casesbracketsshortiii
        {\lambda\, t \,\nainf \nbinf}         {t \ll \tstartwo}
        {\xt^{d+1} \, \nainf \nbinf\, \twid\, t^{(d+1)/z}} 
                                          {\tstartwo \ll t \ll \tl}
        {\xt \, \nainf \, \twid\, t^{1/z}}              {t \gg \tl}
\gap (d<\dc,\  Q>\Qstar)
                                                                \end{eq}
which may be explicitly verified by direct substitution into eq.
\eqref{door}.  For times $t>\tl$, where $\tl \equiv \ta (\nbinf
a^d)^{-z/d}$ is the time to diffuse the mean separation between B
molecules, at least one B lies within the exploration volume of any A
within $\xt$ of the interface.  Hence any such A must have reacted,
and we cross over to 1st order DC kinetics as in eq. \eqref{chair}.
These arguments have implicitly assumed that $\tstartwo<\tl$, \ie
$Q>\Qstar \equiv a h^{-1} \ta^{-1} (\nbinf a^d)^{(d_c-d)/d}$.  For
weakly reactive groups, $Q<\Qstar$, the new 2nd order DC regime is
absent; A reactants collide with many B's before reaction is likely.
The relevant timescale is then $\tstarmany$ and the kinetics of eq.
\eqref{chair} are recovered.

MF theory does not give the correct reaction rate in low dimensions.
In fact the density decay of eq. \eqref{snow} is also incorrect.  For
the symmetric situation, $\nainf=\nbinf$, peculiar correlations
develop at the interface at long times which invalidate this MF decay.
Consider a region of volume $\Omega$, half of which is on the A and
half on the B side.  The fluctuations $\Delta N_\Omega$ in the initial
difference between the number of A and B reactants in $\Omega$ is of
order $(\nainf \Omega)^{1/2}$.  Since reactions conserve this
difference, these difference fluctuations can decay through diffusion
only.  Now if $\Omega \ge \xt^d$, such fluctuations had insufficient
time to decay by $t$.  Hence the density in a region of size $\xt^d$
at the interface is at least $\Delta N_{\xt^d} /\xt^d
\twid t^{-d/(2z)}$.  For $d<\dc$, this is a slower decay than the MF
prediction of eq. \eqref{snow}.  Thus fluctuations determine the
$\nas$ asymptotics in low dimensions:
                                                \begin{eq}{book}
\nas(t) \, \approx\, (\nainf \xt^{-d})^{1/2}\,  \twid\,  t^{-d/(2z)} \comma 
\gap (d<\dc)\period
                                                                \end{eq}
Correspondingly, reactants segregate into A-rich and B-rich regions of
size $\xt$ at the interface.  Such anticorrelations are of course
unaccounted for by the MF approximation, eq. \eqref{approximation}.
These segregation effects are very similar to those found at long
times for bulk 2-species reactions, $A + B \gt
\emptyset$ \citeben{ovchinnikov:segregation_combo}.

\ignore{
We have derived eq. \eqref{book} in a more rigorous way using
arguments which invoked the dynamics of like-particle correlations and
the the assumptions we stated previously.  }

To summarize, we find that an interface lowers the critical dimension,
$\dc=z-1$, relative to simple one-species bulk reactions where $\dc=z$
\citeben{gennes:polreactionsiandii}.  (We note this also is
different to the problem of non-stationary reactive chemical fronts
where for $z=2$, $\dc=2$ has been found
\citeben{galfi:aplusb_front_combo}.)  For spatial dimensions above
$\dc$, densities on either side of the interface are decorrelated and
mean field kinetics apply.  Below $\dc$ strong anticorrelations
develop at the interface.  Correspondingly, a short time 2nd order DC
regime arises for very reactive species, and in the symmetric case at
long times reactants are segregated along the interface and
interfacial densities decay with an anomalous power law in time.  A
peculiarity is that kinetics are of mixed order in the far-field
densities $\nainf, \nbinf$.  Intuition suggests 2nd order kinetics,
since reaction requires an A-B pair to meet at the interface.  But at
long times reaction rates are controlled by diffusion of molecules on
the more dilute A side to the interface, \ie they depend on $\nainf$
only.  The more dense side plays a different role: characteristic
timescales involve $\nbinf$ rather than $\nainf$.

\ignore{ 
The simplest application is small molecules where $z=2$ (Fickian
diffusion) and $\dc=1$.  MF kinetics apply for $d=3$ and $d=2$, while
the one-dimensional case is {\em marginal}.  We have not considered
marginal cases here for reasons of space, but one can show logarithmic
corrections arise for the 2nd order DC regime, $\Rt \twid t/\ln t$.
Numerical testing of this law would be straightforward.
} 

The simplest application is small molecules where $z=2$ (Fickian
diffusion) and $\dc=1$.  MF kinetics apply for $d=3$ and $d=2$, while
$d=1$ is {\em marginal}.  We have not considered marginal cases here
for reasons of space, but we find logarithmic corrections 
to the 2nd order DC regime (the 2nd of the regimes listed in eq.
\eqref{paper}).  The result for $d=1, z=2$ is $\Rt \twid t/\ln t$. 
We have tested our theory for small molecules by numerical simulations
in $d=1$ and $d=2$.  These exhibit 2nd order kinetics  for short times
(see  fig. \ref{reactiface_numerical}(a)), with  MF  kinetics in $d=2$
and logarithmically corrected DC kinetics  in the marginal case $d=1$.
At long times there is a cross-over to 1st order DC behavior with $\Rt
\twid \nainf \,  t^{1/2}$  governed by the  more dilute side; see fig.
\ref{reactiface_numerical}(b).  These numerical results are all
consistent with our theoretical predictions.  On the experimental
side, we hope this work will motivate future studies of, for example,
interfacial polymer systems involving laser-induced macroradicals
\citeben{ben:persistent}.  These can help to to resolve fundamental issues in
interfacial science.

This work was supported by the National Science Foundation, grant no.
DMR-9403566.  We thank Uday Sawhney for helpful discussions.


\pagebreak


\begin{thebibliography}{10}

\bibitem{kotominkuzovkov:book_short}
E.~Kotomin and V.~Kuzovkov, {\it Modern Aspects of Diffusion-Controlled
  Reactions; Cooperative phenomena in Bimolecular processes} (Elsevier,
  Amsterdam, 1996).
\newblock Edited by R. G. Compton and G. Hancock.

\bibitem{ovchinnikov:segregation_combo}
A. A. Ovchinnikov and Ya. B. Zeldovich, Chem. Phys. {\bf 28}, 215-218 (1978);
  D. Toussaint and F. Wilczek, J. Chem. Phys. {\bf 78}, 2642-2647 (1983).

\bibitem{bramson:aplusb_combo_fund}
M. Bramson and J. L. Lebowitz, Phys. Rev. Lett. {\bf 61}, 2397-2400 (1988);
  Phys. Rev. Lett. \ {\bf 62}, 694-694 (1989); Kang K. and Redner S. Phys. Rev.
  A \ {\bf 32}, 435-447 (1985); Zumofen G., Blumen A. and J. Klafter. J. Chem.
  Phys. \ {\bf 82}, 3198-3206 (1985).

\bibitem{gennes:polreactionsiandii}
P.~G. {de Gennes}, J. Chem. Phys. {\bf 76}, 3316--3321, 3322--3326 (1982).

\bibitem{doi:reaction_secondquant1and2}
M.~Doi, J. Phys. A {\bf 9}, 1465, 1479 (1976).

\bibitem{lee:aplusb_rg_all_aip_prl}
B. P. Lee, J. Phys. A {\bf 27}, 2633 (1994); B. P. Lee and J. Cardy, J. Stat.
  Phys. {\bf 80}, 971 (1995); B. Friedman, G. Levine and B. O'Shaughnessy,
  Phys. Rev. A {\bf 46}, R7343 (1992).

\bibitem{doraiswamysharma:book}
L.~K. Doraiswamy and M.~M. Sharma, {\it Heterogeneous Reactions: Analysis,
  Examples, and Reactor Design} (John Wiley \& Sons, New York, 1984), Vol. 1:
  Gas-Solid and Solid-Solid Reactions, Vol. 2: Fluid-Fluid-Solid Reactions.

\bibitem{scott:scottmacosko_gersappe:combo_aip}
C. Scott and C. Macosko, Journal of Polymer Science:Part B {\bf 32}, 205-213
  (1994); D. Gersappe, D. Irvine, A. C. Balazs, Y. Liu, J. Sokolov, M.
  Rafailovich, S. Schwarz,D. G. Peiffer: Science {\bf 265}, 1072-1074 (1994).

\bibitem{galfi:aplusb_front_combo}
L. G\'{a}lfi and Z. R\'{a}cz, Phys. Rev. A {\bf 38}, 3151 (1988); \ S. Cornell,
  M. Droz, Phys. Rev. Lett. \ {\bf 70}, 3824 (1993); \ M. Araujo, H. Larralde,
  S. Havlin, and H. G. Stanley, Phys. Rev. Lett. \ {\bf 71}, 3592 (1993); M.
  Howard and J. Cardy, J. Phys. A \ {\bf 28}, 3599 (1995).

\bibitem{meakinscalapino:catalysis_model}
P.~Meakin and D.~J. Scalapino, J. Chem. Phys {\bf 87}, 731--741 (1987).

\bibitem{lindenberg:aplusb_input}
K.~Lindenberg, B.~J. West, and R.~Kopelman, Phys. Rev. Lett {\bf 60},
  1777--1780 (1988).

\bibitem{chopard:aplusb_semipermeable}
B.~Chopard, M.~Droz, J.~Magnin, and Z.~R\'{a}cz, Phys. Rev. E {\bf 56},
  5343--5350 (1997).

\bibitem{ben:reactiface}
C.~J. Durning and B.~O'Shaughnessy, J. Chem. Phys. {\bf 88}, 7117--7128 (1988).

\bibitem{benfred:reactiface_combo_aip}
B. O'Shaughnessy and U. Sawhney, Phys. Rev. Lett.\ {\bf 76}, 3444 (1996); B.
  O'Shaughnessy and U. Sawhney, Macromolecules \ {\bf 29}, 7230 (1996); G. H.
  Fredrickson, Phys. Rev. Lett. {\bf 76}, 3440 (1996); G. H. Fredrickson and S.
  T. Milner, Macromolecules \ {\bf 29}, 7386 (1996).

\bibitem{ben:persistent}
E.~Karatekin, B.~O'Shaughnessy, and N.~J. Turro, J. Chem. Phys. {\bf 108},
  9577--9585 (1998).

\end{thebibliography}


                     \begin{thefigures}{99}

\figitem{iface_letter}

A and B molecules (size $a$) reacting at an interface of fixed width
$h$ separating immiscible bulks.  Reactions occur within the
interfacial region only.  The local chemical reactivity (rate of
reaction when in contact) is $Q$.  For short times, reactions are
confined to those molecules whose exploration volumes of size $\xt$
overlap at the interface.  The number of such pairs per unit area is
$\xt^{d+1}
\nainf \nbinf$.

\figitem{reactiface_numerical}

A and B random walkers on a square lattice annihilating on contact at
an interface separating A and B bulks with various densities
$\nainf,\nbinf$.  Time in units of site hopping time.  Standard
deviation of mean for each point is less than 3\% in all cases.  Empty
(filled) symbols: $d=1$ ($d=2$).  (a) $\nainf\nbinf t/4\Rt$ v. $t$ for
short times ($\tstarmany,\tl > 10^6$): 2nd order kinetics.  The $d=2$
data asymptote a constant (MF kinetics), while $d=1$ data approach a
straight line, consistent with theoretical law $\Rt\twid t/\ln t$. (b)
$\Rt/\nainf$ v. $t$, long times ($\tstarmany < 100$).  Collapse of data
onto straight line of slope $1/2$ indicates 1st order DC kinetics
governed by dilute A bulk, $\Rt \twid \nainf t^{1/2}$.

\ignore{
\figitem{segregation}

Schematic representation of asymptotic segregation of reactants into A
and B-rich domains of size $\xt$ at the interface, occurring for
$d<\dc$.  This leads to anomalous decay of interfacial densities.
} 

                     \end{thefigures}


\pagebreak

%
%
%

\begin{figure}[t]

\epsfxsize=\textwidth \epsffile{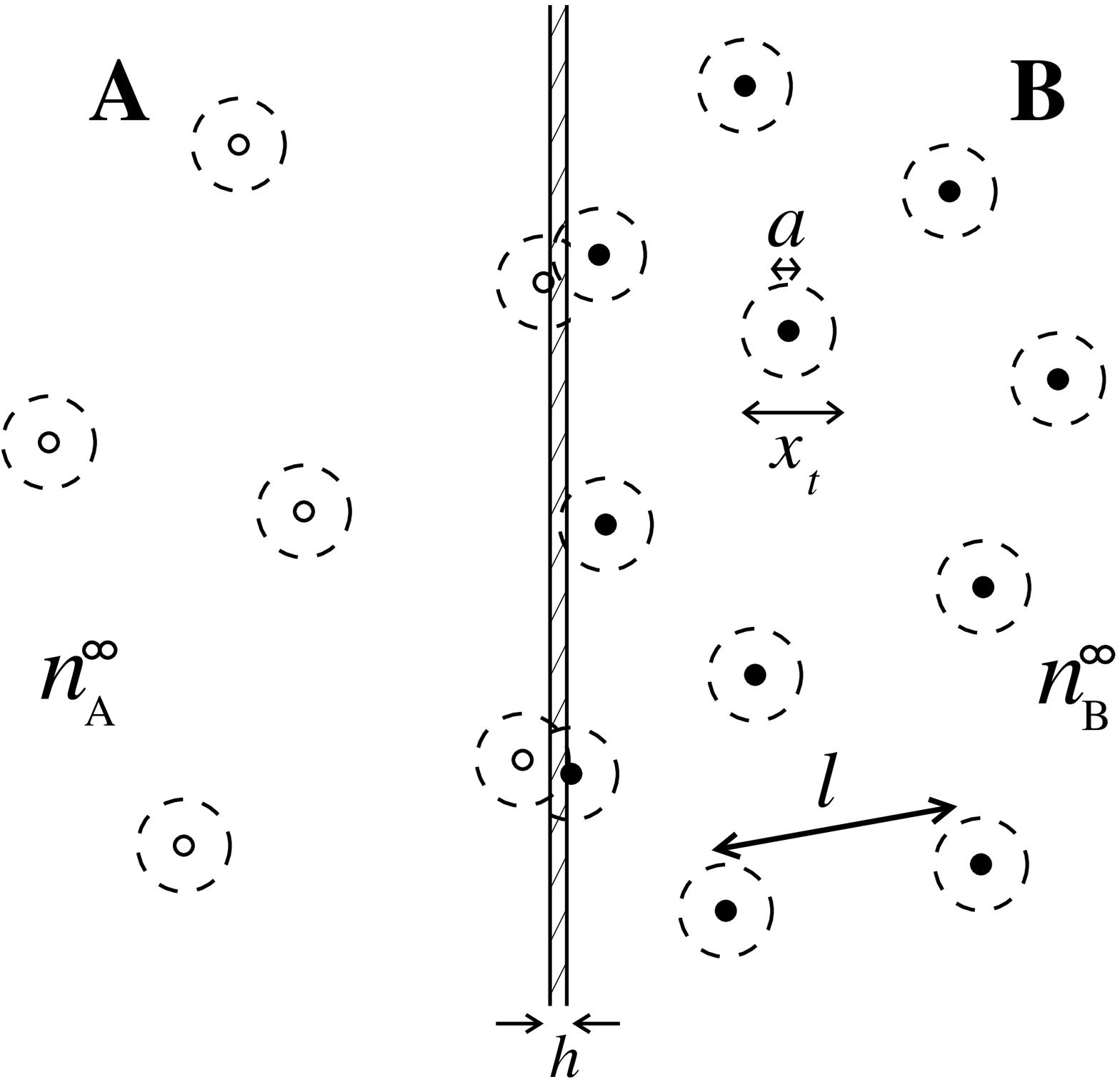}

\end{figure}

\mbox{\ }

\vfill

\addtocounter{fignumber}{1}
\mbox{\ } \hfill {\huge Fig.\@ \thefignumber} 

\pagebreak


\begin{figure}[htb]
\parbox[c]{13cm}{\epsfxsize=13cm \epsffile{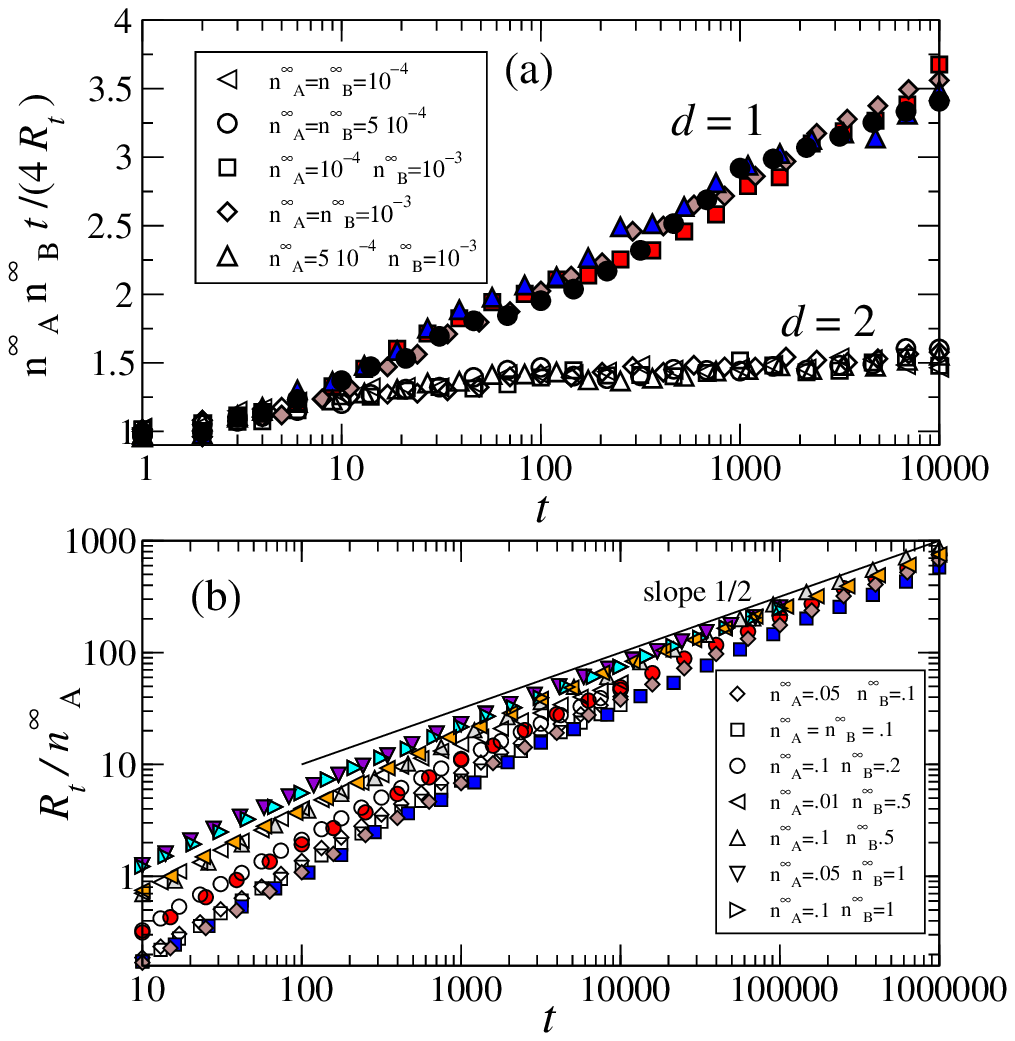}}
\end{figure}

\ignore{
\begin{strip_ft}
{0}
{0}{13}{../figlet2}
{0}
{0}{0}{}
\end{strip_ft}
} 

\addtocounter{fignumber}{1} \mbox{\ } \hfill {\huge Fig.\@ \thefignumber} 

\pagebreak


\end{document}